\documentstyle[12pt,epsf]{article}
\begin{document}
\centerline {EFFECTS OF NONSTOICHIOMETRY IN A$_{3-x}$C$_{60}$ (A= K, RB)}

\vskip0.5cm
\centerline{O. Gunnarsson}             
\vskip0.2cm
\centerline{Max-Planck-Institut f\"ur Festk\"orperforschung,
          Stuttgart, Germany}
\vskip0.3cm
\date{}

\centerline{ABSTRACT}
\vskip0.3cm
We perform self-consistent Hartree    calculations for a model
of A$_{3-x}$C$_{60}$ (A= K, Rb) to study the effects of the
presence of vacancies. We find that the strong vacancy
potential is very efficiently screened and that the density of
states is only weakly influenced by the presence of vacancies.
\vskip0.7cm

\centerline{I. INTRODUCTION}
\vskip0.3cm
The A$_3$C$_{60}$ (A= K, Rb) compounds with the nominal stoichiometry
three are believed to actually have the composition  A$_{3-x}$C$_{60}$,
with $x\sim 0.07$\cite{Stephensx}. The alkali atoms are believed
to be almost completely ionized, and the electrostatic potential from 
the alkali ions should therefore be
large. The presence of a vacancy may then be a  large perturbation
on the electronic structure.
 For instance, if a vacancy were completely unscreened, its potential
at the centre of a neighboring C$_{60}$ molecule would be about 2.3 eV
and it would split the $t_{1u}$ states by about 0.2 eV, which is comparable
to the $t_{1u}$ band width ($\sim 1/2$ eV) for a system without vacancies.
The vacancies occur on the tetrahedral positions\cite{Stephensx} and
each vacancy is surrounded by four C$_{60}$ molecules. If $x=0.07$,
more than a quarter of the C$_{60}$ molecules have an alkali nearest
neighbor vacancy. It is an interesting question how this influences
the electronic structure\cite{RMP}. 
Similar questions arise for the systems Na$_2$Cs$_x$C$_{60}$ and 
K$_{3-x}$Ba$_x$C$_{60}$\cite{Yildirim}, which also have different
charges at the alkali positions.

We perform Hartree    calculations for a model of A$_{3-x}$C$_{60}$.
The model includes the 60 ``radial'' $2p$ orbitals on each C$_{60}$ molecule,
which are allowed to screen the potential from the alkali ions 
self-consistently. We find that the screening is very efficient, in
particular in terms
of charge transfer between the molecules and to a less extent
 in terms of polarizing the molecules.
As a result the density of states is almost unchanged by the presence
of the vacancies.

In Sec. II we present the model, in Sec. III the method for performing
the calculations. The results are given in Sec. IV and some 
implications are discussed in Sec. V.

\vskip0.7cm
\centerline{II. MODEL}
\vskip0.3cm
We introduce a model of A$_{3-x}$C$_{60}$ for which we calculate the
electronic structure and the screening properties.     
For each carbon atom in each molecule we include the $2p$ orbital,
pointing radially out from the molecule. 
The states close to the Fermi energy are mainly formed from these radial 
$2p$ orbitals, and they are therefore particularly important for the
screening. 
The $\sigma$-like $2p$ orbitals pointing
tangentially to the molecular surface and the $2s$ orbitals contribute
mainly to orbitals far away from the Fermi energy and are therefore 
less important. If these orbitals had nevertheless been included, the result
would have been an even more efficient screening than found below. 
Thus we consider the one-particle Hamiltonian 
\begin{equation}
H^0=\sum_{i\nu\sigma}\varepsilon_0 n_{i\nu\sigma}+
\sum_{i\nu j\mu\sigma}\lbrack t(i\nu,j\mu) \psi^{\dagger}_{i\nu\sigma}\psi
_{j\mu\sigma}+ h.c.\rbrack,
\label{eq:10}\end{equation}
where $\varepsilon_0$ gives the energy   of the $2p$ states 
and $t(i\nu,j\mu)$ gives the hopping integrals between these states.
The atoms in a C$_{60}$ molecule are labelled by Roman letters
and the C$_{60}$ molecules are labelled by Greek letters.
This includes both hopping within a molecule and hopping between 
states on nearest neighbor molecules.
We have used the parametrization in Ref. \cite{Orientation},
but multiplied all hopping integrals by a factor 1.2 to obtain
a $t_{1u}$ band width in agreement with recent band structure
calculations.
The C$_{60}$ molecules have an orientational disorder, and take 
essentially randomly one out of two preferred directions\cite{Stephens}.
This is built into the hopping matrix elements\cite{MazinAF}.

The Coulomb integrals between the orbitals are defined as
\begin{equation}
 v(i\nu,j\mu;m\gamma,n\delta)=
\int d^3 r d^3 r^{'} \phi({\bf r}-{\bf R}_{i\nu})
 \phi({\bf r}-{\bf R}_{j\mu   })
{e^2\over|{\bf r}-{\bf r}^{'}| }
 \phi({\bf r}^{'}-{\bf R}_{m\gamma})
 \phi({\bf r}^{'}-{\bf R}_{n\delta}),
\label{eq:11}\end{equation}
where ${\bf R}_{i\nu}$ is the position of the $i$th atom on the $\nu$th
C$_{60}$ molecule.
We neglect the overlap of the functions centred on different
atoms, and therefore $v$ is nonzero only if $(i\nu)=(j\mu)$ and
$(m\gamma)=(n\delta)$.  We make the
assumption
\begin{equation}
v(i\nu,j\mu;m\gamma,n\delta)=
\delta_{i,j}\delta_{\nu,\mu   }
\delta_{m,n}\delta_{\gamma,\delta} \ast
\left\{\begin{array}{ll}
{e^2\over
|{\bf R}_{i\nu}-{\bf R}_{m\gamma}|}
&\mbox{for $|{\bf R}_{i\nu}-{\bf R}_{m\gamma}|>0$}\\
v_0 & \mbox{ for
$|{\bf R}_{i\nu}-{\bf R}_{m\gamma}|=0$}
\end{array}\right.
\label{eq:12}\end{equation}
For the on-site interaction, we have obtained
$v_0=12$ eV from atomic calculations.\cite{trieste}               
For the Coulomb integrals between an alkali core and the carbon 
$2p$ orbitals a similar expression is used.

The C$_{60}$ molecules are put on a fcc lattice with the lattice parameter 
14.24 \AA, and the alkali atoms are placed on the octahedral and tetrahedral 
positions. It was assumed that all C$_{60}$ molecules and
alkali ions are at their ideal positions, and that there are
no distortions due to the vacancies.
 We introduce a large supercell with 500 molecules, 
which is periodically repeated.
This supercell is large enough to give results which are
essentially converged with respect to the cell size.
Within the  supercell each C$_{60}$ molecules is allowed to 
take one of the two preferred orientations randomly.
Furthermore vacancies are introduced randomly on the tetrahedral 
positions in such a way that on the average
there are $3-x$ alkali atoms per unit cell.
To keep the system neutral, we fill up the $t_{1u}$ band with
exactly as many electrons as there are alkali ions in the system.
\vfill\eject

\centerline{III. METHOD}
\vskip0.3cm
We perform self-consistent Hartree calculations. For strongly correlated
systems, this approach might be expected to give poor description
of the screening. By comparing Quantum Monte Carlo and RPA calculations,
it has, however, been found that RPA remains surprisingly accurate
up values of the Coulomb interaction where the Mott-Hubbard
transition takes place\cite{Erik,Yildirim}. 
Since even stoichiometric A$_3$C$_{60}$ is believed to be on the
metallic side of a Mott-Hubbard transition\cite{Mott}, we expect the
Hartree approximation to be qualitatively correct in the present 
situation.

To study the model described above, we use the molecular solid
character of the system. The potential from the
vacancies is screened in two ways. First, there is 
a charge transfer between the molecules to approximately neutralize
the alkali vacancies, i.e., molecules which are nearest neighbors
to a vacancy have a smaller electronic charge.
This screening is very efficient and mainly takes place in the  
partly filled $t_{1u}$ band. Below we only consider the
charge transfer in this band.
 Second, the molecules polarize
so that within a given molecule the charge moves away from a vacancy.
Below we assume that the polarization can be calculated by studying an
isolated molecule in the potential of the surrounding molecules and 
alkali ions.

First we calculate the potential for each atom in each molecule 
due to the the alkali ions and the charges on all other atoms 
in the C$_{60}$ molecules. In this process we take into account that the 
``core'' of a carbon atom has the charge +1, since the $\sigma$ electrons
are counted in the core. 
This calculation is performed using a Madelung summation technique.
For each molecule we then solve the corresponding $60\times 60$ Hamiltonian,
to obtain the $\pi$-orbitals. The lowest 30 orbitals are occupied, 
and the corresponding charges are calculated for the 60 atoms in
each molecule. This describes the polarization of the molecules.

To describe the charge transfer between the molecules, we consider the
three $t_{1u}$ orbitals on each atom. From the molecular calculations
described above, we  obtain the energies of the $t_{1u}$ orbitals,
which in general are not degenerate due to the random       
arrangement of the alkali vacancies. These energies give the diagonal
elements in a $3N\times 3N$ matrix, where $N$ is the number of molecules
in the unit cell. From the calculated coefficients of the $t_{1u}$ 
orbitals and the hopping integrals between $2p$ orbitals on different
molecules, we obtain the hopping matrix elements between $t_{1u}$ 
orbitals on different molecules. These give the nondiagonal
matrix elements in the $3N\times 3N$ Hamiltonian matrix. We use periodic
boundary conditions when setting up the hopping part of the matrix,
which corresponds to considering the ${\bf k}=0$ solution. Due to the
large size of the supercell and the corresponding small size 
of the Brillouin zone, this should be a good approximation. 
Finally, we diagonalize the $3N\times 3N$ matrix and obtain 
the $t_{1u}$ contribution to the charge density.
\vfill\eject 

\begin{figure}[bt]
\unitlength1cm
\begin{minipage}[t]{15.2cm  }
\centerline{\epsfxsize=3.375in \epsffile{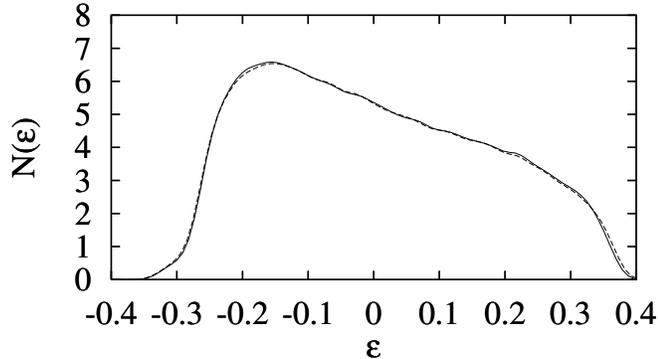}}
\caption[]{\label{fig1} Density of states (per eV-mol-spin)  
for A$_{3-x}$C$_{60}$ for $x=0$ (full curve) and $x=0.07$
(dashed curve). }
\end{minipage}
\hfill
\end{figure}

\centerline{IV. RESULTS}
\vskip0.3cm
The results for the density of states $N(\varepsilon)$ 
are shown in Fig. \ref{fig1}.
Some small wiggles in $N(\varepsilon)$ 
are due to the finite cell size (500 molecules). This does
not, however,  influence the conclusions. The main result of Fig. \ref{fig1}
is that $N(\varepsilon)$  is practically identical for $x=0$
and $x=0.07$. This small difference is due to
the efficient screening of the alkali vacancies. 

To discuss the screening we introduce the average position 
\begin{equation}\label{eq:a1}
\varepsilon_{t_{1u}}(i)={1\over 3}\sum_{m=1}^3   
\varepsilon_{t_{1u}}(im),
\end{equation}
and standard deviation 
\begin{equation}\label{eq:a2}
\sigma_{t_{1u}}(i)^2={1\over 3}\sum_{m=1}^3   
(\varepsilon_{t_{1u}}(im)-\varepsilon_{t_{1u}}(i))^2,
\end{equation}
of the $t_{1u}$ level, where $\varepsilon_{t_{1u}}(im)$ is the energy 
of the $m$th orbital on the $i$th molecule.
We also introduce the average of this standard deviation
\begin{equation}\label{eq:a3}
\sigma_{t_{1u}}={1\over N}\sum_{i=1}^N \sigma(i).
\end{equation}
$\sigma(i)_{t_{1u}}$ and $\sigma_{t_{1u}}$ measure how the 
degeneracy of the $t_{1u}$
orbitals is lifted due to the random positions of the vacancies.
It is also interesting to see how the average $t_{1u}$ position
varies between the sites. For this purpose we introduce the standard
deviation
\begin{equation}\label{eq:a4}
\Delta_{t_{1u}}^2={1\over N}\sum_{i=1}^N (\varepsilon_{t_{1u}}(i)-
\bar \varepsilon_{t_{1u}})^2,
\end{equation}
where $\bar \varepsilon_{t_{1u}}$ is the average of the $t_{1u}$ energies
over the super cell.
In a similar way we define $\Delta_{tetra}$ and $\Delta_{oct}$ as the 
standard deviation in the potentials at the occupied tetrahedra and
octahedral positions, respectively.

As discussed above, there are two types of screening. Due to 
rearrangements of the partly occupied $t_{1u}$ band, charge is 
transferred between the molecules. This influences the averaged
positions $\varepsilon_{t_{1u}}(i)$ of the $t_{1u}$ level 
on the different molecules, and it therefore reduces $\Delta_{t_{1u}}$.
This also reduces the splitting of the $t_{1u}$ levels inside
a molecule. This splitting is further reduced by 
rearrangements of the charge inside the molecules.                    

We therefore perform three types of calculations. First, we perform
a nonself-consistent calculation, where only the potentials from
the alkali ions are considered, and any screening from the C$_{60}$
molecules is neglected. Second, we perform a calculation where
charge transfer between the molecules is allowed, but polarization
of the C$_{60}$ molecules is neglected ($t_{1u}$ self-consistent). 
Finally, we perform the full self-consistent calculation as discussed
above. The results are summarized in Table \ref{tableI}.
 
\noindent
\begin{table}[t]
\caption{On-site splitting $\sigma_{t_{1u}}$ (Eq.~(\ref{eq:a3}))
and off-site splitting
$\Delta_{t_{1u}}$ (Eq.~(\ref{eq:a4})) 
of the $t_{1u}$ level as well as the splitting
in the potential on the tetrahedral ($\Delta_{tetra}$) and octahedral
($\Delta_{oct}$) occupied positions for A$_{3-x}$C$_{60}$. 
The first three lines refer to the system without vacancies ($x=0$)
and the following three lines to a system with $7 \%$ ($x=0.07$). 
The band width $W$ is also shown.
All energies are in eV.}      
\begin{tabular}{ccccccc}
\hline
\hline
                 &$x$ & $\sigma_{t_{1u}}$ & $\Delta_{t_{1u}}$ & 
$\Delta_{tetra}$ &   $\Delta_{oct}$ & $W$  \\
\hline     
Nonself-consistent  &0.00 & 0.00   &  0.00   &  0.00 &0.00& 0.53  \\
$t_{1u}$ self-consistent&0.0  & 0.00   & 0.01    &0.01&0.01 & 0.54 \\
Self-consistent  &0.00 & 0.00   &  0.01   &  0.01 &  0.01 & 0.55  \\
Nonself-consistent  &0.07 & 0.31   &  6.83   & 6.76  & 6.72 & 32.31  \\
$t_{1u}$ self-consistent&0.07  & 0.01   & 0.04  & 0.08& 0.20 & 0.71 \\
Self-consistent  &0.07  & 0.01   &  0.02   &  0.01 & 0.05 &0.55  \\
\hline
\end{tabular}
\label{tableI}
\end{table}

In  the nonself-consistent calculation, we find a very large
variation ($\Delta_{t_{1u}}$) in the average position of the
$t_{1u}$ level between different molecules for $x=0.07$. There is also
a large splitting ($\sigma_{t_{1u}}$) of the $t_{1u}$ level
on molecules close to a vacancy. As a result the band width $W$ is huge. 
When charge transfer 
between the molecules is allowed ($t_{1u}$ self-consistency)
the variation in the $t_{1u}$ level position between the molecules
becomes very small. In the fully self-consistent calculation,
the polarization of the molecules reduces the splitting of 
the $t_{1u}$ level somewhat more. Thus the largest value of 
$\sigma_{t1u}(i)$ is reduced by about a factor of four
and the average value by about a factor of two. Due to the rounding
in Table \ref{tableI}, this is not seen explicitly.
Similar results are found for the potential on the alkali ion
positions, and the potential on the different occupied tetrahedral sites
only varies by a few hundredths of an eV.
\vfill\eject

\centerline{V. CONCLUDING REMARKS}
\vskip0.3cm
We find that 
the strong potential from an alkali vacancy is efficiently screened          
by charge transfer between the molecules and to some extent
also by  polarization 
of the molecules. As a result, the density of states $N(\varepsilon)$
is hardly influenced by the vacancies. 
These results were obtained under the assumption that there are
no distortions of the C$_{60}$ molecules and alkali ions around
the vacancies.

One might have expected that the strong vacancy potential would 
lift the $t_{1u}$ degeneracy on a given molecule, thereby increasing
the band width with vacancy concentration. This would have reduce $N(0)$ and
the electron-phonon interaction $\lambda$, and it might have explained
 the rapid
drop\cite{Yildirim} in $T_c$ as the stoichiometry deviates from three.    
Due to the efficient screening found above, such an explanation,
however, seems unlikely. 

In NMR of Rb$_{3-x}$C$_{60}$ ($x\sim 0.03$) it is found that the 
quadrupolar distortion
of the Rb lines is rather small, in spite of the electric field
gradients one would expect from the vacancies\cite{Mehring}. It was 
argued that this may be due due to an efficient screening 
of the vacancies\cite{Mehring}. Such  considerations agree
very well with the present results. 

\vskip0.7cm
\centerline{ACKNOWLEDGEMENT}
\vskip0.3cm
We would like to thank M. Mehring for helpful discussions.


\begin{thebibliography}{*}

\bibitem{Stephensx}
J.E. Fischer, G. Bendele, R. Dinnebier, P.W. Stephens, C.L. Lin,
N. Bykovets, and Q. Zhu, J. Phys. Chem. Solids {\bf 56}, 1445 (1995);
C.A. Kuntscher, G.M. Bendele, and P.W. Stephens,
Phys. Rev. B {\bf 55}, R3366 (1997). 
  

\bibitem{RMP}O. Gunnarsson, Rev. Mod. Phys. {\bf 69}, 575 (1997).

\bibitem{Yildirim}
T. Yildirim, L. Barbedette, J.E. Fischer, C.L. Lin, J. Robert,
P. Petit, and T.T. M. Palstra, 1996a, Phys. Rev. Lett. {\bf 77}, 167.

\bibitem{Orientation}
O.~Gunnarsson, S. Satpathy, O. Jepsen, and O.K. Andersen, 
Phys.~Rev.~Lett.~{\bf 67}, 3002 (1991).



\bibitem{Stephens}
P.W. Stephens, L. Mihaly, P.L. Lee, R.L. Whetten, S.-M. Huang,
R. Kaner, F. Deiderich, and K. Holczer, Nature {\bf 351}, 632 (1991).


\bibitem{MazinAF} 
I.I.~Mazin, A.I. Liechtenstein, O. Gunnarsson, O.K. Andersen,
V.P. Antropov, and S.E. Burkov,  Phys.~Rev.~Lett.~{\bf 26}, 4142 (1993).

\bibitem{trieste}
O. Gunnarsson, D. Rainer, and G. Zwicknagl, Int. J. Mod. Phys. B
{\bf 6}, 3993 (1992).

\bibitem{Erik} E. Koch, O. Gunnarsson, and R.M. Martin (to be    
publisged).

\bibitem{Mott}O. Gunnarsson, E. Koch, and R.M. Martin, Phys. Rev.
B {\bf 54}, R11026 (1996).

\bibitem{Mehring}
G. Zimmer, K.-F. Thier, M. Mehring, F. Rachdi, and J.E. Fischer,
Phys. Rev. B {\bf 53}, 5620 (1996).



\end{thebibliography}
\end{document}